\documentclass[reprint,a4paper,10pt,twocolumn,aps,prb,nopacs,longbibliography,superscriptaddress]{revtex4-2}
\usepackage[top=1in,bottom=1in,left=0.64in,right=0.64in]{geometry}
\usepackage{csquotes}
\usepackage[utf8]{inputenc}
\usepackage{fancyhdr}
\usepackage{amsmath}
\usepackage{amsfonts}
\usepackage{amssymb}
\usepackage{csquotes}
\usepackage{sidecap}
\usepackage{ragged2e}
\usepackage{subcaption}
\usepackage[nottoc]{tocbibind}
\usepackage{float}
\usepackage{tabularray}
\usepackage[english]{babel}
\usepackage[utf8]{inputenc}
\usepackage{newunicodechar}
\usepackage{caption}
\usepackage{graphicx}
\usepackage{braket}
\usepackage{tikz}
\usepackage{color,soul}
\usepackage[
    colorlinks=True,
    linkcolor=blue,
    filecolor=magenta,
    urlcolor=blue,
    citecolor=blue,]{hyperref}

\begin{document}
\title{Nanoparticle manipulation with a carbon fiber tip in an electron microscope for $\mu$-SQUID magnetometry}
\author{Umesh Chandra Thuwal}
\email{umesht20@iitk.ac.in}
\affiliation{Department of Physics, Indian Institute of Technology Kanpur, Kanpur 208016, India}
\author{Sumanta Maity}
\affiliation{Department of Physics, Indian Institute of Technology Kanpur, Kanpur 208016, India}
\author{Clemens B. Winkelmann}
\affiliation{\mbox{Univ.} Grenoble Alpes, CNRS, Grenoble INP, IRIG-PHELIQS, Grenoble 38000, France}
\author{Hervé Courtois}
\affiliation{\mbox{Univ.} Grenoble Alpes, CNRS, Grenoble INP, Institut Néel, Grenoble 38000, France}
\author{Anjan Kumar Gupta}
\email{anjankg@iitk.ac.in}
\affiliation{Department of Physics, Indian Institute of Technology Kanpur, Kanpur 208016, India}

\date{\today}

\begin{abstract}
We report a carbon-fiber-tip based nanomanipulation system integrated into a scanning electron microscope for individual nanoparticle (NP) manipulation on a surface. Electrochemically etched amorphous carbon fiber tips with excellent mechanical rigidity and sub-100 nm apex radii effectively reduce the van der Waals adhesion and enable reliable positioning of about 100 nm size NPs with about 100 nm precision. This system combines a piezoelectric bimorph for vertical tip motion, a four-quadrant piezo-tube for two-dimensional fine tip control and a two-dimensional piezoelectric walker for coarse lateral translation. Using this setup, we successfully position single Fe$_3$O$_4$ magnetic NPs on micron sized superconducting quantum interference devices for optimal magnetic coupling between them and probe a NP's magnetism.
\end{abstract}

\maketitle

\section{Introduction}
Nanoparticles (NPs), particularly magnetic NPs, exhibit physical properties that differ significantly from their bulk counterparts due to effects such as quantum confinement \cite{wilson1993quantum}, surface-dominated interactions, and finite-size behavior \cite{khan2019nanoparticles,vazquez2011finite}. Individual NP studies enable the observation of discrete and anisotropic magnetic switchings due to magneto-crystalline anisotropy \cite{jalili2019effect}, quantum tunneling \cite{wernsdorfer2007classical} of magnetization, super-paramagnetism \cite{mikhaylova2004superparamagnetism}, and stochastic relaxation dynamics \cite{brown1963thermal}. These phenomena are critical for emerging technologies, including spintronics, where single magnetic nano-particles can function as local memories \cite{wolf2010promise}, and biomedical applications such as targeted drug delivery and magnetic hyperthermia for cancer treatment \cite{xiao2020superparamagnetic,Farzin2020}. Furthermore, magnetic NPs are promising for high-density data storage systems \cite{wolf2010promise}.

To probe an individual NP using small sensors such as micron-sized superconducting quantum interference devices ($\mu$-SQUIDs) or magnetic tunnel junctions, a NP must be precisely positioned near such detectors to enable optimal coupling. Accurate placement also minimizes undesired interactions with other particles and ensures a controlled local environment, essential for reliable measurements \cite{wernsdorfer2007classical, wernsdorfer2009micro}. However, achieving such precise placement presents several experimental challenges. A wide range of micro- and nano-manipulation strategies have been developed for NP manipulation over past three decades with each strategy being tailored to different particle sizes, geometries, and device environments \cite{junno1995controlled,xie2009versatile, bartels1997basic, Kim2011,Shah2024, Mazerolle2005, Ye2013, Molhave2006, Cao2017, Liu2019, Bartenwerfer2011,  Dong2007, Shi2016,Liu2019D,Zai2025A,Zai2025B}. Mechanical surface-force-based methods, including atomic force microscope (AFM), have been used for manipulation of NPs as small as 10–20 nm \cite{junno1995controlled,Kim2011,Burger2022}. The AFM operation relies on raster imaging, which is not real time, and offers limited three-dimensional control. It is also susceptible to drift-induced positioning errors, tip deformation, and unintended particle pickup. The latter becomes more common with NP size reducing below 20 nm as adhesion dominates \cite{Kim2011,Shah2024,argento1996parametric} over inertial and other forces. Other strategies have therefore been investigated, including field-driven repositioning and indirect AFM manipulation \cite{Kim2011,Liu2019D}.

Real-time-image guided NP manipulation inside a scanning electron microscope (SEM) by ragging of NP or picking-and-placing using tungsten probes or using dual-probe systems or MEMS microgrippers \cite{Mazerolle2005,Ye2013,Molhave2006,Cao2017,Liu2019} has also been quite successful. These approaches, often assisted by electron beam induced deposition (EBID), have enabled automated transfer and placement of nano-sized objects. Building on these capabilities, such systems have evolved into nanorobotic platforms featuring piezoelectric multi-axis actuation, sub-$\mu$m positioning resolution, and closed loop visual-servo control \cite{Shah2024,Bartenwerfer2011,Dong2007,Shi2016}. Despite substantial progress, most gripper systems remain unreliable for sub-micron NPs due to strong adhesion and uncontrolled release. At these scales, van der Waals (vdWs) forces dominate tip–particle interactions which complicates reliable dragging \cite{junno1995controlled}. While sharp tip apex reduces adhesion by minimizing contact area \cite{argento1996parametric}, mechanical instability of ultrasharp metallic tips often limits manipulation. Effective NP positioning therefore requires probes that combine nanoscale sharpness, mechanical robustness and real-time visual feedback and control \cite{lee2023effects}.

In this paper, we present a nanopositioning system for manipulating individual NPs inside an SEM. A carbon fiber tip with large elastic limit and a nanoscale apex is used for NP manipulation. Operating inside an SEM provides real-time visual feedback with about 100 nm NP placement precision. We successfully placed Fe$_3$O$_4$ NPs of sizes 150~nm or larger on niobium $\mu$-SQUIDs to achieve optimal magnetic coupling. An Fe$_3$O$_4$ NP's magnetic reversal through vortex nucleation and annihilation was studied with a $\mu$-SQUID at 2.5 K temperature.

\section{Tip considerations}
The tip plays a critical role in NP manipulation. A sharp tip with mechanically stable tip-apex together with overall rigidity and large elastic limit is essential for reliable nano-manipulation. Tungsten (W) tips, although widely used and easy to fabricate, are mechanically brittle, prone to fracture at small radii \cite{ekvall1999preparation} and their high surface energy leads to strong vdW adhesion with NPs \cite{junno1995controlled,argento1996parametric}. In addition, tungsten’s low elastic limit results in poor mechanical resilience under nanoscale loading \cite{French2000}.

Carbon fiber is an excellent candidate as it possesses large elastic limit, low surface energy and high rigidity. Electrochemically etched carbon fibers also yield sharp apex radii ($<100$~nm), yet mechanically robust tips for precise and reproducible NP manipulation. Another critical factor in tip fabrication is its cone angle and taper length. A large taper length with small cone angle makes the tip more fragile. Thus, tips with not so small cone angle but small apex radii are most desirable.

The magnitude of vdW adhesion force, between the NP and the tip, as well as between the NP and substrate, plays an important role in manipulation. In all cases, gravity is negligible and electrostatic forces, such as those induced by the SEM beam, can be larger but still vdW forces dominate below a certain NP size. The vdW force depends on the NP’s shape, size, and its contact area with the tip and the substrate. Within a simple sphere–sphere contact model, the contact radius scales as $a_0\propto R_{\rm eff}^{2/3}$ \cite{lyu2021automated,yuk2015shape} such that an increased tip radius leads to a substantially larger contact area. Here, $R_{\rm eff}\approx R_{\rm t}R_{\rm NP}/(R_{\rm t}+R_{\rm NP}$) with $R_{\rm t}$ and $R_{\rm NP}$ as the tip and NP radii, respectively.

For spherical tip and particle in contact, the Hamaker model gives  $F_{\rm vdW}=A_{\rm H} R_{\rm eff}/6d^2$ \cite{hamaker1937london}. Here, $A_H$ is the Hamaker constant which depends on the material and the medium and $d$ is the surface-to-surface distance. A larger, blunter tip yields stronger adhesion, which makes the detachment of NP from the tip harder. In contrast, a small apex tip will reduce the $F_{\rm vdW}$. For example, a 10 nm radius feature amounts to less than 1 nN vdW force. This assumes an $A_{\rm H}=10^{-19}$ N-m and a $d=5$ $\AA$ Thus, sharp tips with small apex radii weaken the adhesion, making it easier to manipulate NPs. Additionally, the NP shape, surface roughness, its atomic facet structure and surfactants at the surface will play a role in dictating the tip-NP interaction force.

\section{Carbon fiber tip optimization}
\begin{figure}[hbt!]
	\centering
	\includegraphics[width=3.29in]{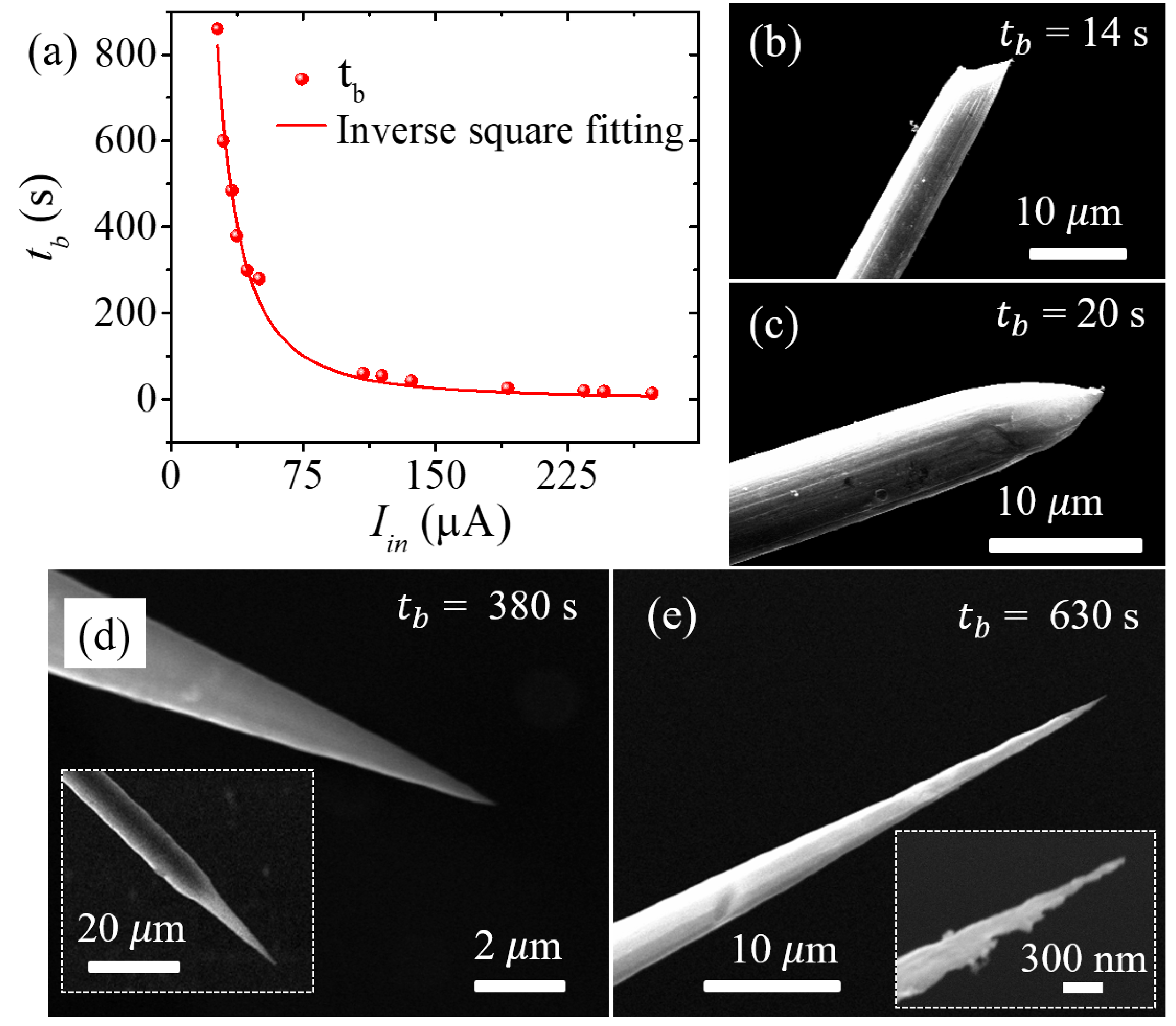}
	\caption{(a) Breaking time $t_{\rm b}$ variation with initial current $I_{in}$ and an inverse-square fitting as given by Eq. \ref{eq:inv-sq}. (b) and (c) show the tips etched at $I_{\rm in}=272$ and 236~$\mu$A, respectively. (d) Tip etched at $I_{\rm in}=38$~$\mu$A with t$_b$ = 380~s, yielding a well-tapered apex of $\lesssim 100$~nm suitable for manipulation. The inset shows a larger-scale view. (e) Tip etched at $I_{\rm in}=39$~$\mu$A with $t_{\rm b}=630$~s produces an ultrasharp, but fragile, tip of apex radius close to 35~nm.}
	\label{fig:fig1}
\end{figure}
Cleaved \cite{Zai2025A} or etched \cite{Gomez2010,Meza2015} carbon fiber tips have been used in manipulation and scanning probe microscopy imaging. We used an electrochemical etching method to fabricate sharp tips of amorphous carbon fibers \cite{C-fiber-source} of approximately 6 $\mu$m diameter fixed to a 0.25 mm diameter tungsten wire as described in the Suppl-Info \cite{suppl-info}. The sharpness of the tip is found to be dictated by the initial etching current $I_{\rm in}$ that determines the breakage time of the tip $t_{\rm b}$ at which the etching finishes and the tip breaks at the air–electrolyte interface. Figure ~\ref{fig:fig1}(a) shows $t_{\rm b}$ as a function of $I_{\rm in}$. Remarkably, this $t_{\rm b}$ dependence can be fitted to an inverse-square relation, see Fig.~\ref{fig:fig1}(a), given by,
\begin{align}
t_{\rm b} = \frac{K}{I_{\rm in}^{2}}.
\label{eq:inv-sq}
\end{align}
The fitted value of $K$ was found to be $0.57\times10^{-6}$ A$^2$.s. The etching time can be expected to depend on the total amount of positive charge $Q_{\rm p}$ that flows from the tip to the electrolyte. This $Q_{\rm p}$, governing the amount of carbon etched by oxidation, will be proportional to the time-integral of the current magnitude. If we assume a linear reduction, to zero, in current with time, the time $t_{\rm b}$ required for a fixed $Q_{\rm p}$, in order to etch a given amount of carbon, will be proportional to $1/I_{\rm in}$. From Fig. \ref{fig:fig1}(b-e), the taper-length of the tip is seen to increase with increasing $t_{\rm b}$ or reducing $I_{\rm in}$. This implies an increase in the amount of etched carbon with decreasing $I_{\rm in}$ and thus, a sharper rise in $t_{\rm b}$, with decreasing $I_{\rm in}$, than just $1/I_{\rm in}$. Note that this is only a qualitative argument. In detail, the instantaneous etching rate and the etched profile will depend on the diffusion of species in the vicinity of the tip. This requires detailed modeling and further investigation.

Figures~\ref{fig:fig1}(b-e) show the changes in tip shape when $t_{\rm b}$ changes from 14 s, corresponding to $I_{\rm in}=272$ $\mu$A, to 630 s, for $I_{\rm in}=29$ $\mu$A. For $t_{\rm b}=14$ s, the tip is quite blunt. It becomes conical at $t_{\rm b}=20$ s and has the most desirable shape at $t_{\rm b}=380$ s. For $t_{\rm b}=630$ s (at $I_{\rm in}=29$ $\mu$A), in Fig. \ref{fig:fig1}(e), the tip has a long tapered end with a very small cone angle and about 35 nm apex radius, which is close to the smallest achievable value. However, such ultrasharp tips were found to be unsuitable due to their mechanical fragility.

In summary, for NP manipulation in our setup, a $t_{\rm b}$ range of 350-400~s, corresponding to an $I_{\rm in}$ range of 35-40~$\mu$A, is found to be optimum. The resulting tips exhibit an apex radius $\lesssim 100$~nm and they are not so fragile. The tip shown in Fig. \ref{fig:fig1}(d) was successfully used for positioning an Fe$_3$O$_4$ NP on a $\mu$-SQUID. $t_{\rm b} < 350$ s can be good for larger size particles while $t_{\rm b} > 400$ s could be useful for smaller NPs provided the tip is protected from excess mechanical stress.

\begin{figure}[ht]
	\centering
	\includegraphics[width=3.29in]{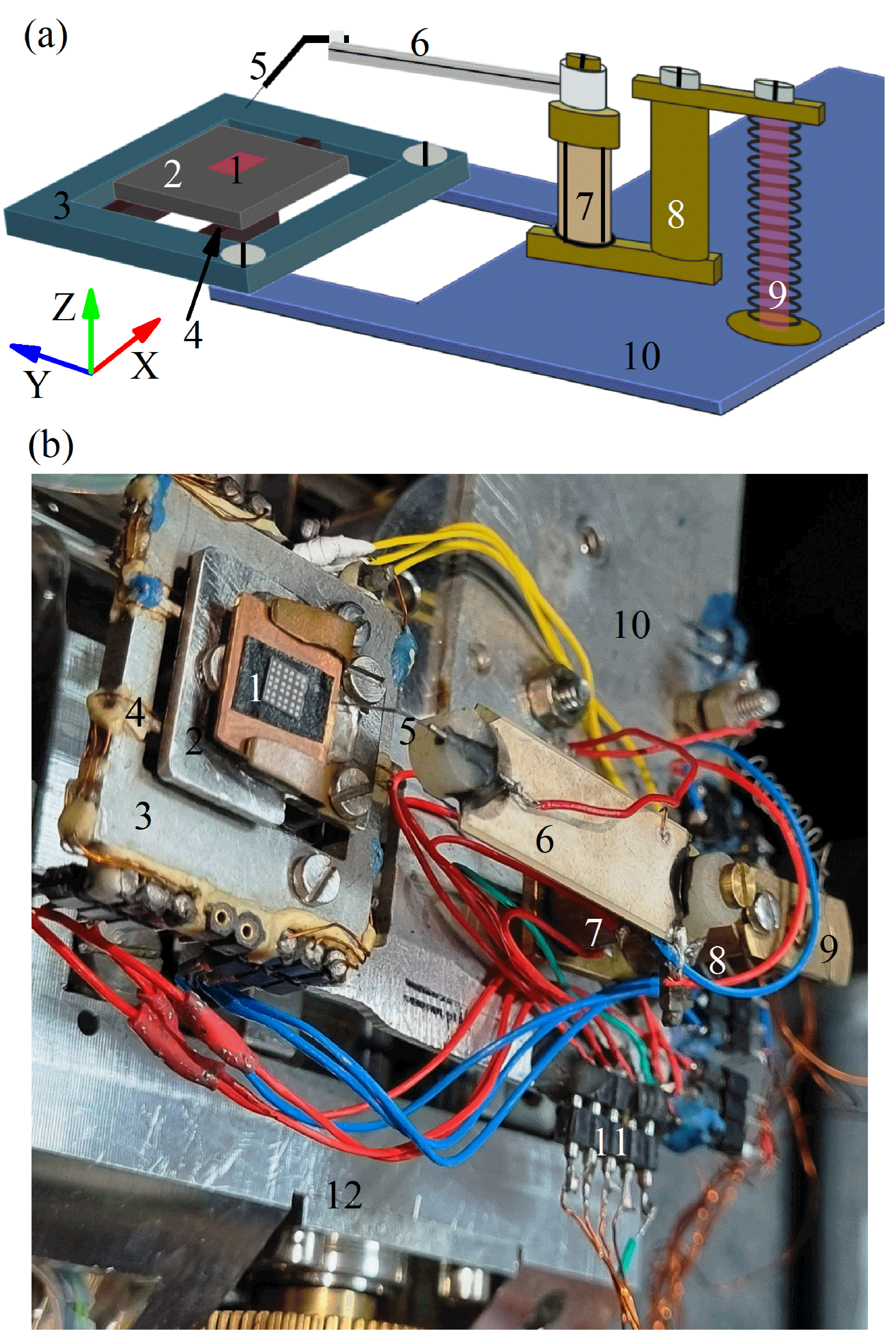}
	\caption{(a) Simplified schematic of the nano-manipulator system with following components: sample (1), 2D walker platform (2), SS frame of 2D walker (3),  PE bimorph actuators of 2D walker (4), tungsten wire with the carbon fiber tip (5), PE bimorph for z-motion of tip (6), PE tube (7), brass rod (8), spring loaded stage with setscrew (9) and aluminum platform (10). (b) Complete nano-manipulation assembly mounted inside the SEM with connecting wires (11) and SEM stage (12). In (b) one can make out the array of nine $\mu$-SQUIDs, on the sample (1), with each SQUID having four contact pads.}
    \label{fig:fig2}
\end{figure}
\section{Nano-manipulator design}
Figure \ref{fig:fig2} shows the nanomanipulator system, which can be divided into two parts, namely, the tip holder assembly and the 2D walker which are rigidly fixed on an aluminum platform. Further zoomed-in pictures of the nanomanipulator are given in the Suppl-Info \cite{suppl-info}. The tip holder assembly consists of a spring-loaded stage, a piezoelectric tube (PE tube) and a piezoelectric $Z$-bimorph. The $Z$-bimorph is used for fine tip motion in the $Z$-direction with about 10 nm precision using a DC-voltage and this motion is limited to about 100 $\mu$m. The bimorph is fixed on a PE tube \cite{binnig1986single} that has its outer electrode segmented into four quadrants and a common inner electrode. This tube provides fine movement, in the $XY$-plane, of the tip attached to the bimorph. This $XY$ movement is restricted to a few microns but has nm precision. The spring-loaded stage, see Figs. \ref{fig:fig2}(a) and (b), with a fine-pitch setscrew is used to bring the tip within the range of the $Z$-bimorph. A stainless steel capillary is fixed to the free end of this $Z$-bimorph which is used to hold the W wire with etched carbon-fiber tip attached to its other end by Ag paint.

\begin{figure*}[ht]
	\centering
	\includegraphics[width=6.75in]{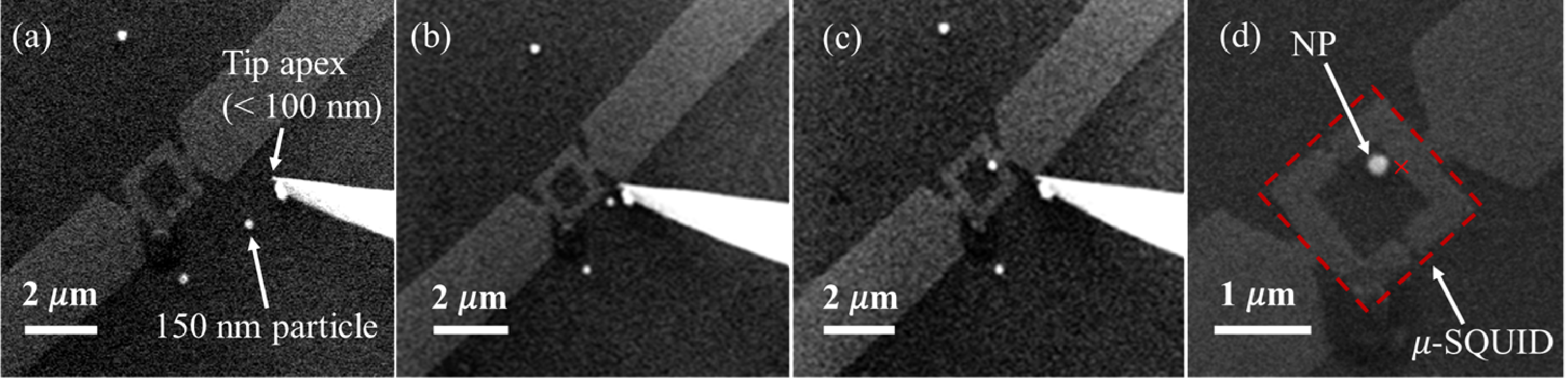}
	\caption{Nano-particle manipulation using a sharp tip with an apex radius less than 100 nm. (a) The particle, approximately 150 nm in size, is initially located about 2 $\mu$m away from the $\mu$-SQUID. (b) The particle is successfully dragged using the tip and brought closer to the $\mu$-SQUID, at a distance of about 500 nm. (c) The NP is successfully placed in the middle of the top right inner edge of the $\mu$-SQUID loop. (d) A closeup view of the $\mu$-SQUID with the NP in the final state. The outer edge of the SQUID loop has been marked by red discontinuous line. The NP is slightly shifted by about 100 nm from its previous position (marked by the cross) of image (c) to its current position. Note that the overall contrast in these SEM images, taken during manipulation, is dictated by the visibility of all the three entities, namely, the $\mu$-SQUID, NPs and the tip. This results into a relatively poor contrast of the $\mu$-SQUID.}
	\label{fig:fig4}
\end{figure*}
The 2D walker \cite{gupta2008compact} has about 200 nm step size and it can move the sample stage up to about 1 mm in both $X$-$Y$ directions. The dual-actuation architecture provides enhanced freedom with fine control between tip and substrate in all the three directions. The sample substrate, with NPs dispersed on its top surface, is mounted on this 2D walker platform. The tip-holding capillary and the sample platform are electrically grounded to the aluminum frame. This ensures zero potential difference between tip and sample and avoids electrostatic interactions. The manipulator assembly is high-vacuum compatible, permitting operation under SEM for real-time visual feedback.

\section{Placement of ${\rm Fe}_3{\rm O}_4$ NP on $\mu$-SQUID}
The niobium film $\mu$-SQUIDs are fabricated on a silicon substrate using electron beam lithography \cite{paul2020probing}. Fe$_3$O$_4$ NPs are synthesized using a hydrothermal method \cite{geng2008controllable} and characterized using X-ray diffraction and high resolution transmission electron microscope as presented in the Suppl-Info \cite{suppl-info}. The NPs are dispersed by sonication in n-hexane together with oleic acid and oleylamine as surfactants to avoid NP agglomeration. A silicon chip with an array of $\mu$-SQUIDs is dipped into the sonication bath for a few seconds to enable a random distribution of NPs on the chip. This chip is dried and mounted on the 2D-walker of the nano-manipulator. The carbon-fiber tip is mounted on the $Z$-bimorph so that it makes about $45^{\circ}$ angle with the chip-surface. This serves three purposes: 1) the tip and the substrate are simultaneously visible from top in the SEM, 2) it facilitates pushing of the NP from a side by the tip and, 3) it reduces the force on the tip during its first contact with the substrate due to fiber's elastic deformation.

To operate the nanomanipulator, the tip is first brought in close proximity, visible with bare eyes, of the substrate using the set-screw of the spring-loaded stage and than the manipulator assembly is inserted into the SEM. The tip is gradually approached towards the substrate using the $Z$-bimorph which enables the tip's gentle contact with the substrate. This is carried out under the SEM's monitoring with the tip kept in SEM focus. The sample stage is then translated using the 2D walker to bring a selected NP into the tip's vicinity. During any large lateral translation, the tip is slightly retracted to prevent its damage and adhesion of NPs to it. Once the desired particle is positioned close to the tip, the substrate is carefully translated using 2D walker to bring the target $\mu$-SQUID close to the stationary tip, which remains in the SEM field of view as it drags the NP on the moving substrate. This permits a NP placement with sub-$\mu$m precision. For more precise positioning, in the final stages, the PE tube provides the required fine lateral movement.

Figure~\ref{fig:fig4} shows the SEM images taken during the manipulation of a NP where an Fe$_3$O$_4$ NP is dragged by several microns and positioned on the $\mu$-SQUID. In Fig.~\ref{fig:fig4}(a), the labeled NP, together with two other NPs, is seen at about 2 $\mu$m distance from the $\mu$-SQUID edge. In this image, One can also see two NPs adhering to the side of the tip. This is found to happen occasionally but such attachment is typically seen to occur away from the apex due a larger contact area. However, the tip apex remains sharp and clean, permitting successful manipulation of a NP. The $\mu$-SQUID loop with two parallel constrictions is marked in Fig.~\ref{fig:fig4}(d). In Fig.~\ref{fig:fig4}(b), the NP is moved close the right top corner of the SQUID while in Fig.~\ref{fig:fig4}(c), the NP has been placed at an inner edge of the SQUID. This NP displacement was done using 2D walker while in the final stage, see Fig.~\ref{fig:fig4}(d), the NP is shifted with, sub 100nm precision, to slightly away from the inner edge of the SQUID and using the fine tip control.

\section{Magnetic reversal study of an ${\rm Fe}_3{\rm O}_4$ NP}
The sensitivity of a $\mu$-SQUID to a NP's magnetic moment is inversely proportional to SQUID's linear size. It also depends on the location of the NP relative to the SQUID loop. A good coupling between the two is achieved when the NP is placed close to the inner edge of the SQUID loop and the maximum coupling is achieved for a NP on the $\mu$-SQUID's constriction. The NP's magnetic signal needs to be above the SQUID's noise-limited sensitivity and it is desired that it is within the SQUID's linear and monotonic sensitivity range. Thus, a small NP needs to be more strongly coupled as compared to a big one. Note that a SQUID's response to magnetic-flux is sinusoidally periodic with period $\Phi_0 \approx 2\times 10^{-15}$ T.m$^2$ rather than monotonic. For the $\mu$-SQUIDs used here, the effective loop area is 1.67 $\mu$m$^2$ so that a 1.2 mT perpendicular field amounts to a $\Phi_0$ flux. The limitation due to large NP flux can be partially overcome by the feedback mode of the SQUID. In this mode, a small perpendicular magnetic field $B_{\rm f}$ compensates the NP flux by keeping the total SQUID flux fixed using its voltage signal. However, due to its limited speed the feedback will not be able to track very large and abrupt flux jumps, due to NP's magnetic response, that exceed $\Phi_0/2$. In fact, for the NP in Fig.~\ref{fig:fig4}, this was found to be the case due to a very strong coupling.

\begin{figure}[ht]
	\centering
	\includegraphics[width=3.23in]{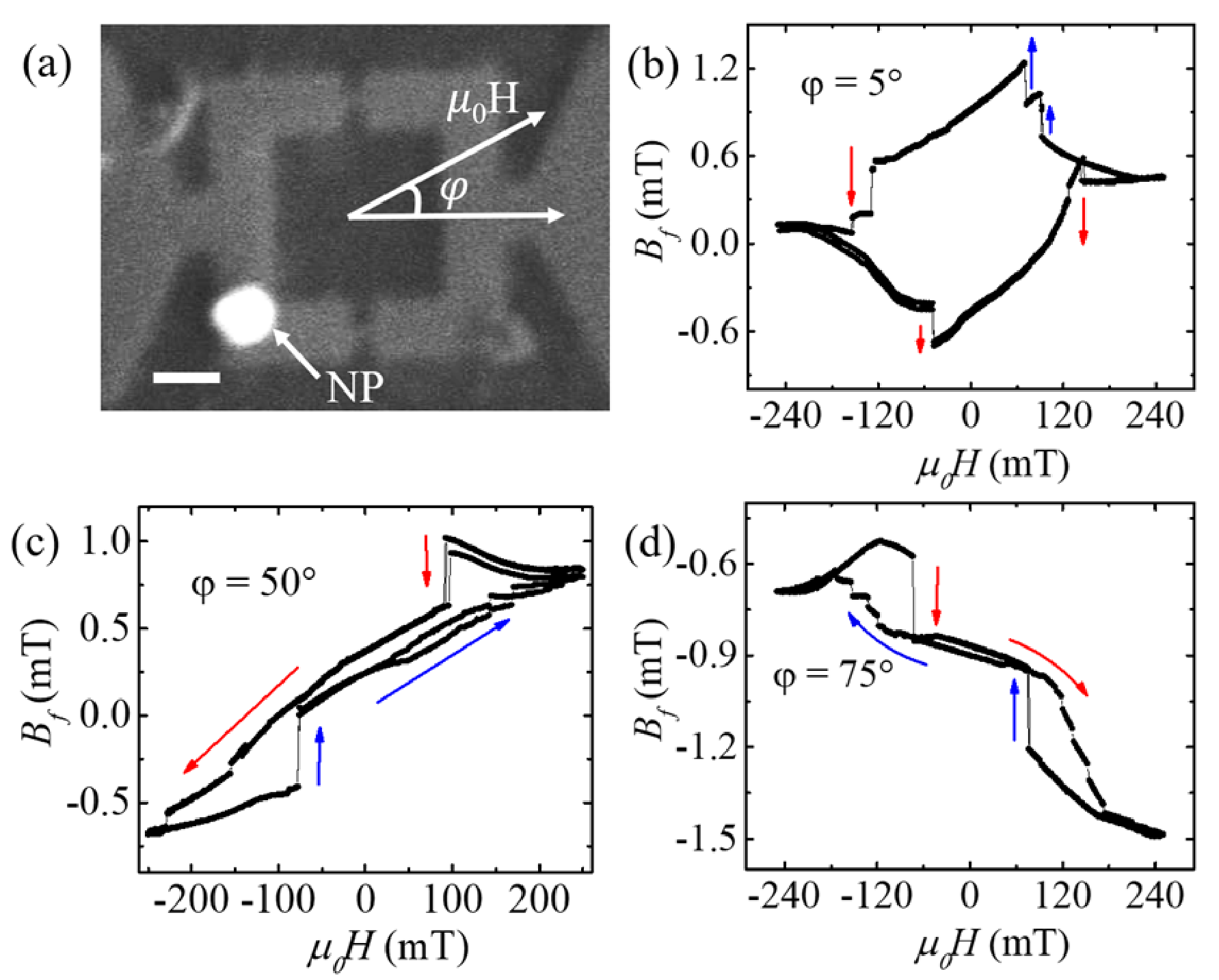}
	\caption{(a) Fe$_3$O$_4$ NP positioned on the $\mu$-SQUID for magnetic reversal measurements with in-plane magnetic field $\mu_0 H$ applied at different angles $\varphi$ relative to the SQUID. Magnetic hysteresis loops of the NP at (b) $\varphi=5^\circ, (c) 50^\circ$ and (d) $75^\circ$. These loops exhibit multiple abrupt jumps together with a continuous change as the magnetic vortex or domain wall traverses through the NP.}
	\label{fig:fig5}
\end{figure}

Figure \ref{fig:fig5} shows the magnetic characteristics of another single Fe$_3$O$_4$ NP. Here, the Fe$_3$O$_4$ NP is positioned just outside a corner the $\mu$-SQUID, see Fig. \ref{fig:fig5}(a). These measurements were carried out in a liquid helium based setup with a superconducting vector magnet. This setup and measurement details are described elsewhere \cite{paul2020probing}. The magnetic field is applied within the SQUID plane so that the SQUID only responds to the NP's magnetic moment. The field angle $\varphi$ within the SQUID-plane as defined in Fig. \ref{fig:fig5}(a) is varied to probe the anisotropic magnetic reversal in this NP. These measurements were done using the feedback mode as this NP's overall signal exceeds the $B_{\rm f}$ range of 1.2 mT, corresponding to $\Phi_0$, though all the abrupt jumps are well below $\Phi_0$/2. The $B_{\rm f}$–H loops, measured within a magnetic field range of –250 mT to +250 mT, in Fig. \ref{fig:fig5}(b-d) at three different $\varphi$ values, show a clear anisotropy. This arises from the shape and magnetocrystalline anisotropy of this Fe$_3$O$_4$ NP. Since this NP size exceeds the single domain limit, the reversal happens through inhomogeneous magnetization modes involving vortex or domain wall nucleation, traversal and annihilation \cite{paul2022magnetization}.

\section{Discussion and Conclusions}
Our nano-manipulation technique, using multiple stages of piezoelectric actuation, real-time SEM imaging, and optimum carbon fiber tips, enables positioning of single NPs with about 100 nm precision. While the need and advantages of the present technique were discussed in the Introduction section, below we discuss some of the operational aspects and limitations. For a successful manipulation, the adhesion force of the NP with the surface should dominate over that with the tip. The contact area of a NP on the side of the tip can be comparable to that with the substrate below the NP. Thus, to avoid NP adhesion to the tip, NP should only be touched by the apex of the tip, and not by its side. This aspect will depend on the NP size as well as its shape. The manipulation of NPs of size less than 100 nm by this method appears difficult at present. Our tip-etching results show that getting a sharper tip is not the bottleneck but mechanical fragility of a very sharp tip is a problem. A more careful protocol on tip handling during operation or some other protection method may help in using very sharp tips and manipulating smaller NPs successfully. In this sense, the operator skill and experience has some role in successful manipulation and a systematic statistical optimization remains a subject for future work.

Finally, the successful measurement of anisotropic magnetic hysteresis in a 200 nm sized NP placed on a $\mu$-SQUID validates this technique. The ability to resolve magnetic signals from individual NPs underscores both the exceptional flux sensitivity of the $\mu$-SQUID and the precision of the NP placement technique. The nano-manipulator-assisted positioning can ensure optimum magnetic coupling between the NP and the SQUID loop. Beyond its immediate relevance for fundamental studies of nanoscale magnetism, the demonstrated approach offers promising technological advantages for scalable integration in quantum sensing, spintronic devices, and high-resolution nano-magnetic characterization.

\section*{Acknowledgements}
AKG acknowledges funding through SERB Project CRG/2021/000273 and financial support from IIT Kanpur.


\begin{thebibliography}
\bibitem{C-fiber-source} The amorphous carbon fibers of this type are manufactured by several companies. We tried these from three different sources, including two from Amazon, that show similar results.
\end{thebibliography}
\end{document}